\documentclass[journal]{IEEEtran}
\usepackage[english]{babel}      
\usepackage[utf8]{inputenc}   
\usepackage{graphics}
\usepackage{subfigure}
\usepackage{graphicx}
\usepackage{epsfig}
\usepackage[centertags]{amsmath}
\usepackage{graphicx,indentfirst,amsmath,amsfonts,amssymb,amsthm,newlfont}
\usepackage{longtable}
\usepackage{cite}
\usepackage[usenames,dvipsnames]{color}
\usepackage{amsmath,amsfonts,amssymb,amsthm,epsfig,epstopdf,titling,url,array}
\usepackage{hyperref}
\newcommand{\doi}{doi:}
\renewcommand*{\doi}[1]{\href{http://dx.doi.org/#1}{doi: #1}}

\ifCLASSINFOpdf
\else
\fi
\begin{document}
%
\title{An Overview of Self-Similar Traffic:\\ Its Implications in the Network Design}
%
%
%

\author{Ernande F. Melo~and~H. M. de Oliveira
\thanks{Ernande F. Melo is with Department of Computer Engineering, State University of Amazonas (UEA), Av. Darcy Vargas, 1200, Parque 10, 69065-020, Manaus, Amazonas, Brazill.}
\thanks{H. M. de Oliveira is with Departament of Statistics, Federal University of Pernambuco (UFPE), Recife, Brazil.}}
\maketitle
%
%
%
\begin{abstract}
The knowledge about the true nature of the traffic in computer networking is a key requirement in the design of such networks. The phenomenon of self-similarity is a characteristic of the traffic of current client/server packet networks in LAN/ environments dominated by network technologies such as Ethernet and the TCP/IP protocol stack. The development of networks traffic simulators, which take into account this attribute, is necessary for a more realistic description the traffic on these networks and their use in the design of resources (contention elements)  and protocols of flow control and network congestion. In this scenario it is recommended do not adopt standard traffic models of the Poisson type.
\end{abstract}


%
\IEEEpeerreviewmaketitle

\section{Introduction}
%
%
%
%
\IEEEPARstart{T}he recurrent use of computer networks advances with notorious growth. Network applications have multiplied, and, particularly, social networks do not stop growing up. This high-growing is mainly due to the popularization of the use of computers, mobile phones and the easy access to the Web. The transmission of data of audio and video through such networks presents demands that vary over time and requires differentiated services (high-speed data, channel band reservation and priority traffic to audio or video) generating traffic with high speed and great variability. Knowledge about the very nature and behavior of the network traffic emerges as a fundamental requirement for the design, operation and maintenance of networks, demanding the execution of several activities, among which one can highlight
\begin{itemize}
\item
Sizing of resources: fundamental activity of design and maintenance. The dimensioning of resources aims to qualify and quantify the network components installed. Typical sizing examples
are the communication channel and the allocation of contention-node buffers, such as \textit{routers} and \textit{switches}.
\item
Quality of service (QoS): maintaining the network quality of service is always a challenge and predicting the peak usage remains crucial for deciding which corrective measures can be adopted.
\item
Protocol Design: As a basis for network operation, the adequacy of protocols to the type of network traffic is essential. In this case, designing and implementing new policies for flow control and congestion is a key part of the process.
\end{itemize}

\begin{figure}[!ht]
\begin{center}
\includegraphics[width=3.5in]{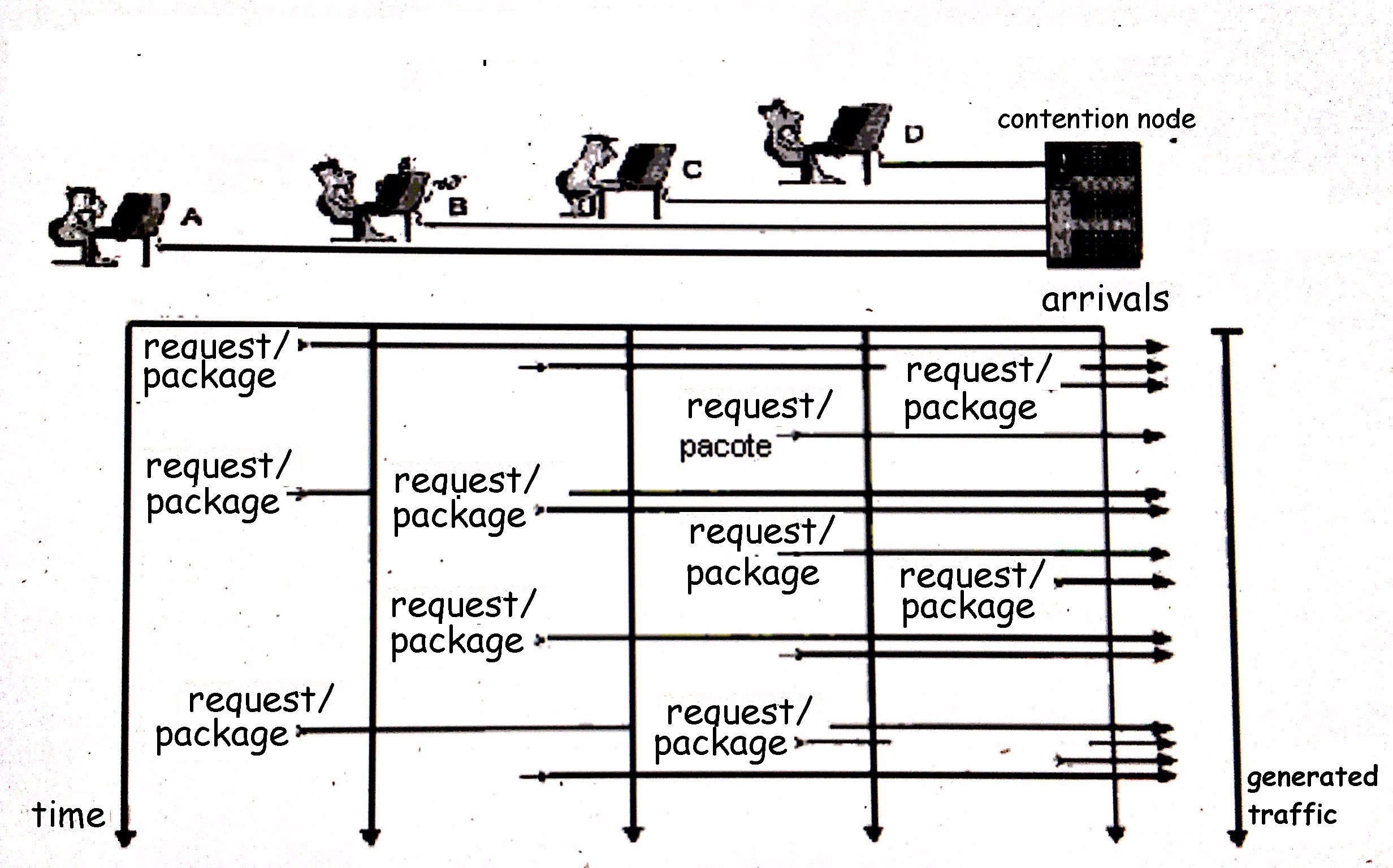}
\end{center}
\caption{Representation of incoming traffic in a containment node due to requests/packages generated by stations. The generated traffic is described by an arrival process over time, in a containment node.}
\label{fig:fig1}
\end{figure}

\begin{figure}[!ht]
\begin{center}
\includegraphics[width=3.5in]{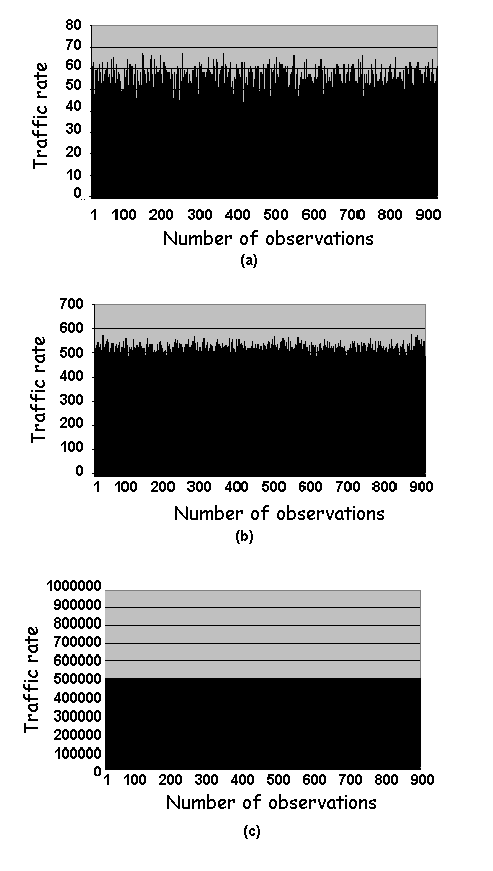}
\end{center}
\caption{SRD case: Representation of simulated traffic with ``visual'' SRD characteristic observed at intervals of: (a) 1 unit of scale (b) 10 unit of scale  (c) 10,000 unit of scale.  The maximum number of observations in each of the three sub-figures (axis of abscissa) is different in each case, using 900, 9,000 and 9,000,000 observations, respectively.}
\label{fig:fig2}
\end{figure}

In this sense, the implementation of network traffic generators more closely of real networks traffic is essential to the design, operation and maintenance activities. These processes of traffic generation are based on observation of the number of requests/packets arriving at containment nodes of the network in a given time interval. A sequence of such observations represents the generated traffic. Figure~\ref{fig:fig1} illustrates such a process. It illustrates the different moments in which each of the four different users place their respective requests for transmitting packets. On the right side of the figure, the arrows indicate the traffic demand generated jointly by users at the containment node.
 The most widely spread network traffic models (both for analysis as generation) still use Poisson processes, or more generally, Markovian processes, due to their simplicity, with consequent abridgement in the traffic model \cite{Frost_Melamed}. These models were inherited of the voice traffic models of telephone networks. The main characteristics of this models are:
\begin{enumerate}
\item
Arrivals of requests/packets are random events;
\item
Arrivals of requests/packages are independent events, i.e., the arrival of one package is not determinant for another to arrive; property known as \textit{memoryless};
\item
In the analysis, the counting interval is divided into sub-intervals  $(\Delta \to 0)$ so small that at most one or no event is observed;
\item
Distribution of the number of requests/packets is proportional to the duration of the observed time interval, i.e., the arrival process follows a pre-determined rate.
\item
The probability distribution for number of the arrival of packages is discrete.
\end{enumerate}
The characteristics of the Poisson process can be illustrated in  Figure ~\ref{fig:fig2}, considering the users (A, B, C, D) transmitting randomly and independently only one arrival/packet request each time (in each case, time $\Delta \to 0)$; the mean number of arrivals is proportional to the time of observation). Among the characteristics observed in the Poisson process, the most relevant to the traffic model based on this process is related to its memoryless property, which means that the current traffic behavior is independent of the previous traffic history. This implies in predicting that the traffic, when observed into increasing intervals of time scales (from milliseconds to minutes and even hours) tends to smooth in a few scales of observation time. In other words, the peaks rapidly dissolve with increasing time scale of observation. Statistically, this means that the structure of autocorrelation, which measures the degree of dependence between current traffic and previous traffic, decays, undoes rapidly and that this traffic has  short dependence interval or Short-Range Dependence (SRD). Figure ~\ref{fig:fig2} shows a sequence of graphs illustrating (in a very simplified way) a traffic generated with this characteristic in three scales (1 unit of reference scale, 10 units of scale and 10,000 scale units). In the sequence of the graphs, the peaks are undoing with the increase of the scale of observation time scale. These graphics were generated by means of a Pareto probability distribution function, to be described in the sequence of this work, with parameters conveniently adjusted to illustrate the SRD characteristics. Baccelli \cite{Baccelli} argue that until the 1990’s, modeling the traffic by traditional statistical laws was still adequate in many cases. But traffic then changed and it was necessary to understand how and why. This model started to be questioned from 1993 onwards, results from the work of Leland, Taqqui, Willinger and Wilson \cite{Leland_et_alli}, on traffic in Ethernet LAN and followed by other works related to packet traffic, namely: Ethernet Local Network  \cite{Leland_et_alli2}, Wide Area Network (WAN) \cite{Paxson_Floyds}, \cite{Crovella_Bestavros}, Variable bit rate video (VBR) over Asynchronous Transmission Mode (ATM) \cite{Beran_et_alli}, CCSN/SS7 \cite{Duffy_et_alli} and Integrated Services Digital Network (ISDN) \cite{Erramilli_Willinger}. These works and others \cite{Willinger_et_alli} \cite{Crovella_Bestavros2} showed that, contrary to the Poisson traffic model, the actual traffic of these networks, when observed on multiple timescales, does not  smooth  -- instead, it displays peaks across multiple scales of observation time, when compared to traffic with SRD characteristics. Statistically this means that the traffic autocorrelation structure is maintained for several time scales and the traffic presents a characteristic of long dependence interval or \textit{Long-Range Dependence} (LRD) \cite{Hosking}. Figure ~\ref{fig:fig3} combines a sequence of graphs illustrating (simplified mode) traffic generated with this characteristic, a three scales (the same ones used for the graphs in  Figure~\ref{fig:fig2}). It can be seen from the graphs that the peaks are not redo with the increasing scale of observation time. The graphs were generated by means of a distribution function of Pareto probabilities, with parameters conveniently adjusted to illustrate the LRD characteristic.

The LRD characteristic is relevant in sense that the traffic behavior, when observed through different time scales, is of great importance in the performance analysis, the flow control and network congestion, since the demand for buffers allocation in nodes of containment, latency of actuation of flow control mechanisms and congestion and, to some extent, the demand for bandwidth, depend on the sampling time scale or traffic observation. This directly implies questioning about the validity of results obtained in performance analysis and network control using Poisson-based models of traffic, that is, models with SRD characteristics used as base for the activities of network design, operation and maintenance. 

\section{Self-similar Traffic Model}

\begin{figure}[!ht]
\begin{center}
\includegraphics[width=3.3in]{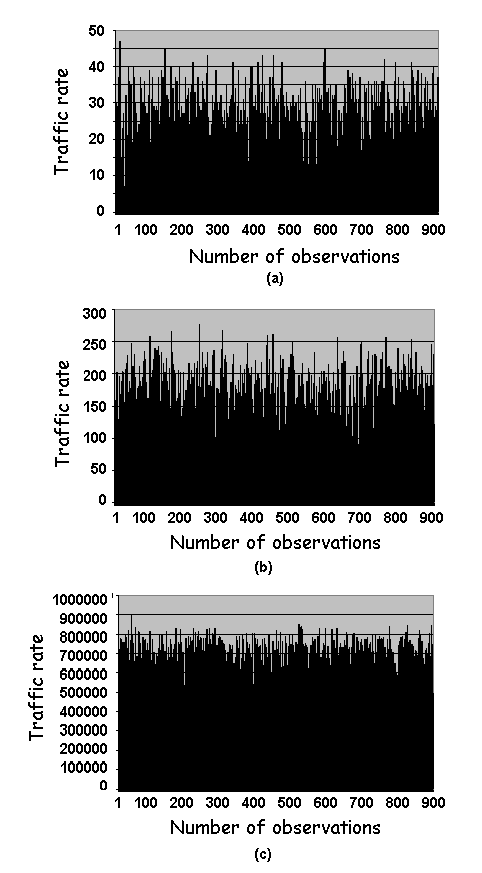}
\end{center}
\caption{LRD case: Representation of simulated traffic with ``visual'' LRD characteristic observed at intervals of: (a) 1 unit of scale (b) 10 unit of scale  (c) 10,000 unit of scale.  The maximum number of observations in each of the three sub-figures (axis of abscissa) is different in each case, using 900, 9,000 and 9,000,000 observations, respectively. Nevertheless, the overall traffic  profile does not change substantially.}
\label{fig:fig3}
\end{figure}

Self-similarity experienced a growing explosion in publications and multiple applications in the 1990s and 2000s \cite{Leland_et_alli}, \cite{Erramilli_Willinger}, \cite{Jeong Hae-Duck_et_alli}, \cite{Feldmann_et_alli1},  \cite{Kihong_et_alli}, \cite{Lopes-Ardao_et_alli}. The search for a traffic model that included the characteristics LRD led researchers in the field to associate this characteristic with the idea of ``fractals'' that are objects that maintains the same appearance when observed at different scales  \cite{Peitgen_et_alli}, \cite{Mandelbrot_Wallis}, \cite{Mandelbrot}, \cite{Ryu}. Looking at Fig. ~\ref{fig:fig3}, some self-similar behavior can be observed by inspection, by noting that the traffic profile remains rather the same at different time scales.  This is in contrast with the traffic shown previously, in Fig. ~\ref{fig:fig2}, which represents traffic with SRD characteristics. The processes used to describe this fractal nature of network traffic are called self-similar processes \cite{Peitgen_et_alli}, \cite{Mandelbrot_Wallis}. These present structural similarity through a large number of observation scales, that is, the structure repeats at multiple scales and in different contexts or different dimensions (geometric, statistical or temporal) and the behavior of the process statistics, such as mean, variance and the autocorrelation function does not change with spatial or temporal scale variation (stationary processes) \cite{Granger_Joyeux},\cite{Bruce_Disney}, \cite{McDysan}, \cite{Kihon_Walter},  \cite{Mandelbrot}. In this work, a brief description of self-similar traffic is made and its implications in network design investigated.

\subsection{Model of traffic in local networks}

\begin{figure}[!ht]
\begin{center}
\includegraphics[width=3.5in]{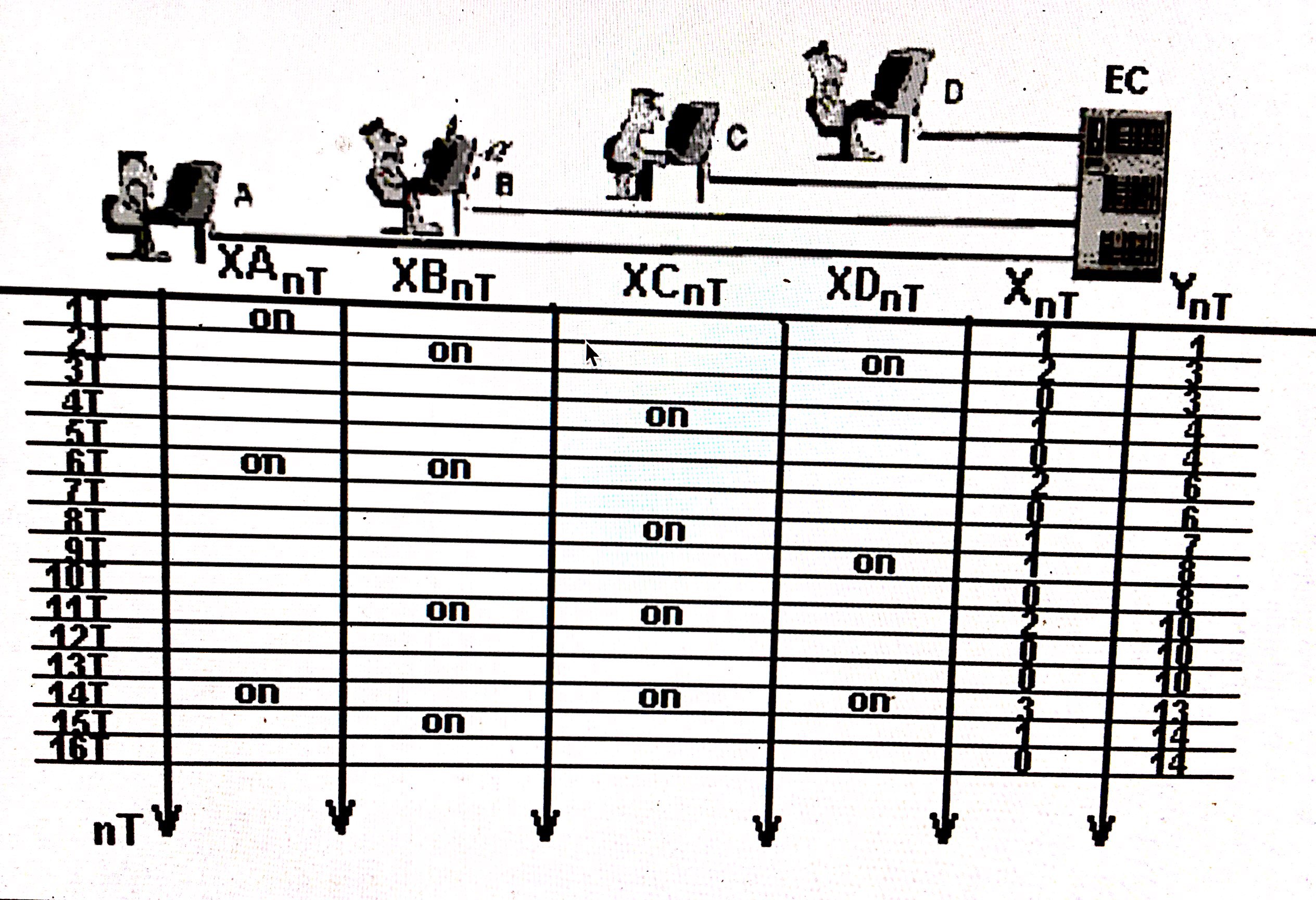}
\end{center}
\caption{Network environment showing 4 stations (say A, B, C, D) transmitting ($on$ state) on a containment element (CE) at intervals time (\textit{slots}) $nT$. Each of signals $\{X^A_{n}\},~\{X^B_{n}\},~\{X^C_{n}\}$ e $\{X^D_{n}\}$ are stochastic processes that represent the amount of bits or packets transmitted, at each time interval $nT$. Here, $\{X_{n}\}$ represents the incremental process related to the aggregate or accumulated traffic of the four stations, arriving at the containment element CE (arrival process), in each time interval $nT$. The variable $Y_{n}$ represents the cumulative process, referring to traffic accumulated from time zero to the nth interval $nT$.}
\label{fig:fig4}
\end{figure}

Traffic in a local network has two components. One referring to the sources of traffic (stations and servers) which, by generating traffic in a random manner, can have their traffic modeled as a stochastic process \cite{Bruce_Disney}. The other component refers to the traffic itself that can be seen as an arrival of packets in a network contention element, switch or router, that can be represented as a time series, in which the application of statistical methods as aggregation, measures (mean, variance, autocorrelation) and spectral analysis, determines whether the process is self-similar. Figure~\ref{fig:fig4} illustrates a process of generation traffic, in which A, B, C and D represent the sources of  generating traffic in a random way, at discrete time $T$ steps $(X^A_{n}, X^B_{n}, X^C_{n}, X^D_{n})$, EC represents the element of contention, $X_{n}$ represents the incremental process and $Y_{n}$ represents the cumulative process, related to cumulative traffic from time zero to $n$-th interval $nT$. In this figure, the demands of each source are exemplified, in 16 consecutive time intervals ($1T$ to $16T$). The purpose of this illustration is to clarify the \textit{counting process} that results in the variables of the incremental process $X$ and the cumulative process $Y$. See, for example, that in the time slot corresponding to $11T$, there are only two sources acting (namely, B and C), so that $X11=2$ (column XnT). The accumulated values for the demand correspond to a sum with 11 terms: $1 + 2 + 0 + 1 + 0 + 2 + 0 + 1 + 1 + 0 + 2 = 10$ (column YnT). For the sake of simplicity, one can assume normalized instants setting $T=1$. The random traffic of generating sources can be modeled by a probability distribution that, for the current networks should be able to capture the high variability in traffic network. The most appropriate distribution is the Pareto distribution of the long tail (\textit{heavy-tailed}) \cite{Paxson_Floyds}, \cite{Crovella_Bestavros}. 

\subsection{Characterization of Self-Similar Processes}

A self-similar process that captures the LRD-nature is characterized by a single parameter, called \textit{Hurst parameter} ($H$) \cite{Averill_David}, \cite{Leland_et_alli}, \cite{Lapponi} in a range  $0.5\leq H<1$, being the process classified with a low degree of self-similarity when $H$ is close to 0.5 and with high degrees of self-similarity when it is close to 1. 
The value of $H$ can be estimated from three different approaches. (i) Time domain analysis based on R/S statistics \cite{Leland_et_alli2}, \cite{Beran}, (ii) Analysis of variances of the aggregate process $\{Y^m_n\}_n$, in which $m$ is the aggregation level $m \in \mathbb{Z}^+$ \cite{Leland_et_alli2} and (iii) domain analysis frequency (periodogram) \cite{Beran}. 

Rescaled adjusted range statistics (or R/S method) is a graphical method for estimating hurst parameter of self-similar network traffic \cite{Gospodinov-Gospodinova}, \cite{Lenskiy-Seol}. An aggregate traffic stream is a collection of flows that are grouped together for common treatment between two points in a network. All aggregate packets are subject to the same traffic management policies. The value of $H$ is related to the slope measurement of line in a log $\times$ log-chart. In this work both methods (Aggregate Variance and R/S) are used owing to their implementation simplicity, with a good level of precision.
 
\begin{figure}[!ht]
\begin{center}
\includegraphics[width=3.5in]{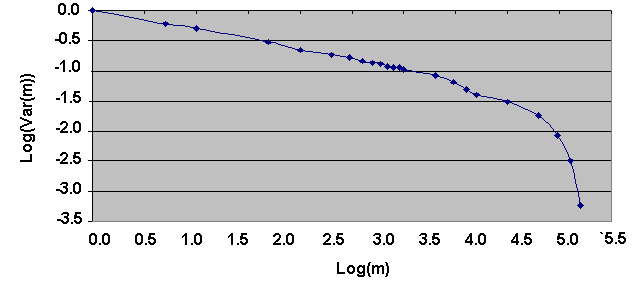}
\end{center}
\caption{Application of the graphical method of Variance Aggregate for a sub series of data collected by Bellcore Morris Research and Engineering Center (MER). The parameter $m$ is the selected aggregation level. The estimated slope ($\beta$) is 0.39, for the interval between cut-off points 1.0 and 4.0, corresponding to an estimate of $H=1-\beta /2=0.8$.}
\label{fig:fig5}
\end{figure}

For the aggregate variance approach, the parameter $H$ is measured by $H=1-\beta/2$, in which $\beta$ is the slope of the graph line $\log(Variance\_Aggregate(m) \times \log(m)$. The graph of Figure ~\ref{fig:fig5} represents the application of the method of Aggregate Variance to evaluate the degree of self-similarity of a set of data collected by \textit{Bellcore Morris Research and Engineering Center} (MER). The series corresponds to traffic values one hour of normal use of an Ethernet network, collected in units of time of 10 milliseconds, resulting in a sample of size $n= 360,000$. The values of the series represent the number of packets per unit of time. These data were first analyzed in \cite{Leland_et_alli},\cite{Leland_et_alli2}. The result
obtained for the value of the parameter $H$ was $H=0.8$, while the value obtained in this work was from $H \approx 0.8$, the inclination being estimated at 0.39 for the interval between cut-off points 1.0 and 4.0.

\begin{figure}[!ht]
\begin{center}
\includegraphics[width=3.5in]{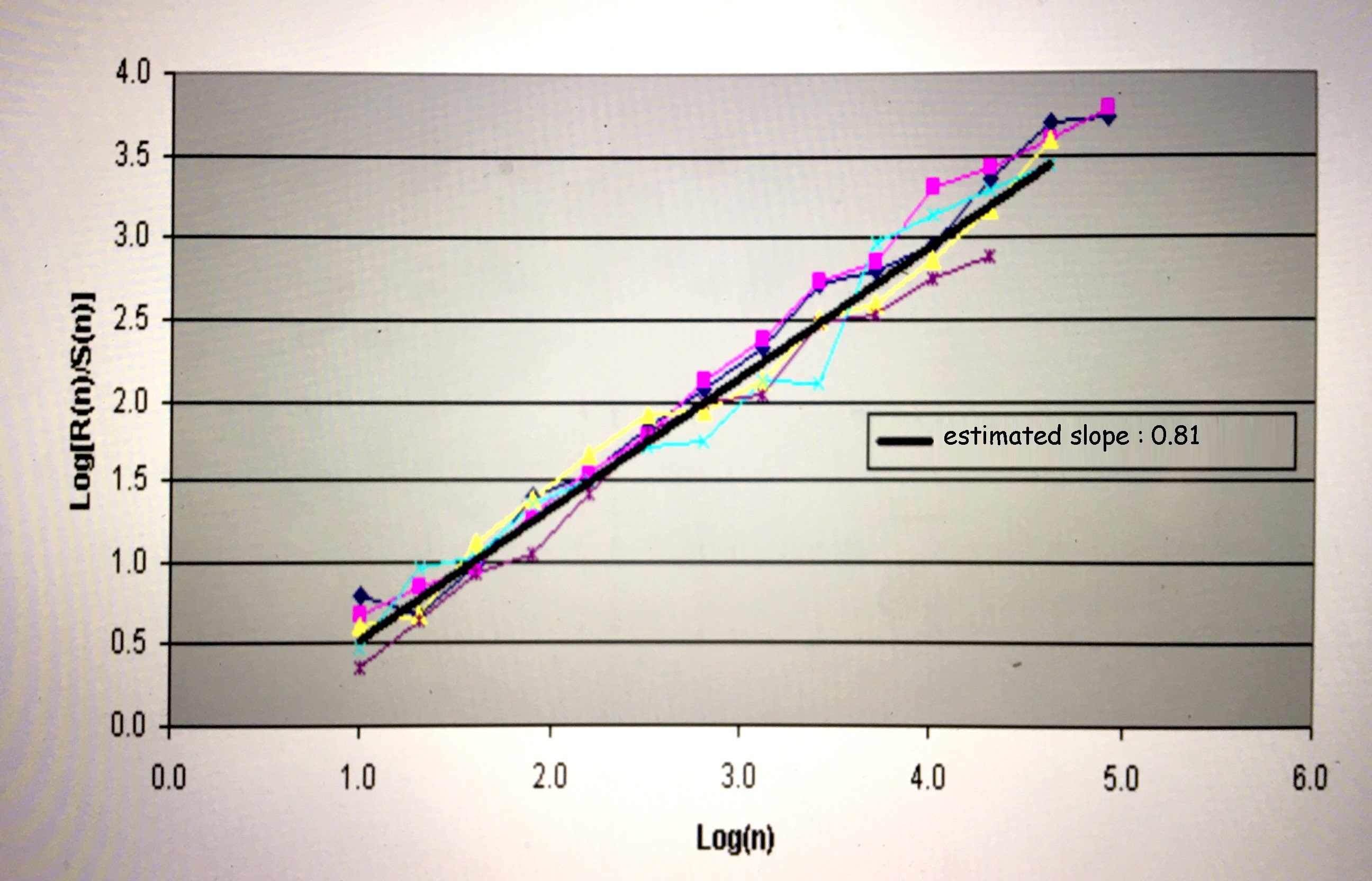}
\end{center}
\caption{Illustration of the R/S graphical method for estimating Hurst parameter for the series that partially represents the data set by Bellcore Morris Research and Engineering Center (MER). The estimated slope, which corresponds to the value of $H$, is 0.81.}
\label{fig:fig6}
\end{figure}

For the R/S method the parameter $H$ is measured directly as the slope of the line in the graph $\log(R(n)/S(n) \times \log(n)$). Figure~\ref{fig:fig6} represents the application of the method R/S to evaluate the degree of self-similarity for the same set of data from the previous scenario. The estimate obtained for parameter $H$ was $H=0.81$, which is perfectly in line with the range of typical values expected for a traffic of LRD-nature.

\subsection{Model for Self-Similar Traffic Generation}
Among the models for self-similar traffic generation, one can cite the \textit{Fractional Brownian Motion} (FBM), the \textit{Fractional Gaussian Noise} (FGN) and the \textit{On/Off} model. The focus is a model that captures the essential elements of a local network, namely independent sources generating traffic that is aggregated in a medium shared by that sources. In this case, the \textit{On/Off} model is the most suitable.

\begin{figure}[!ht]
\begin{center}
\includegraphics[width=3.5in]{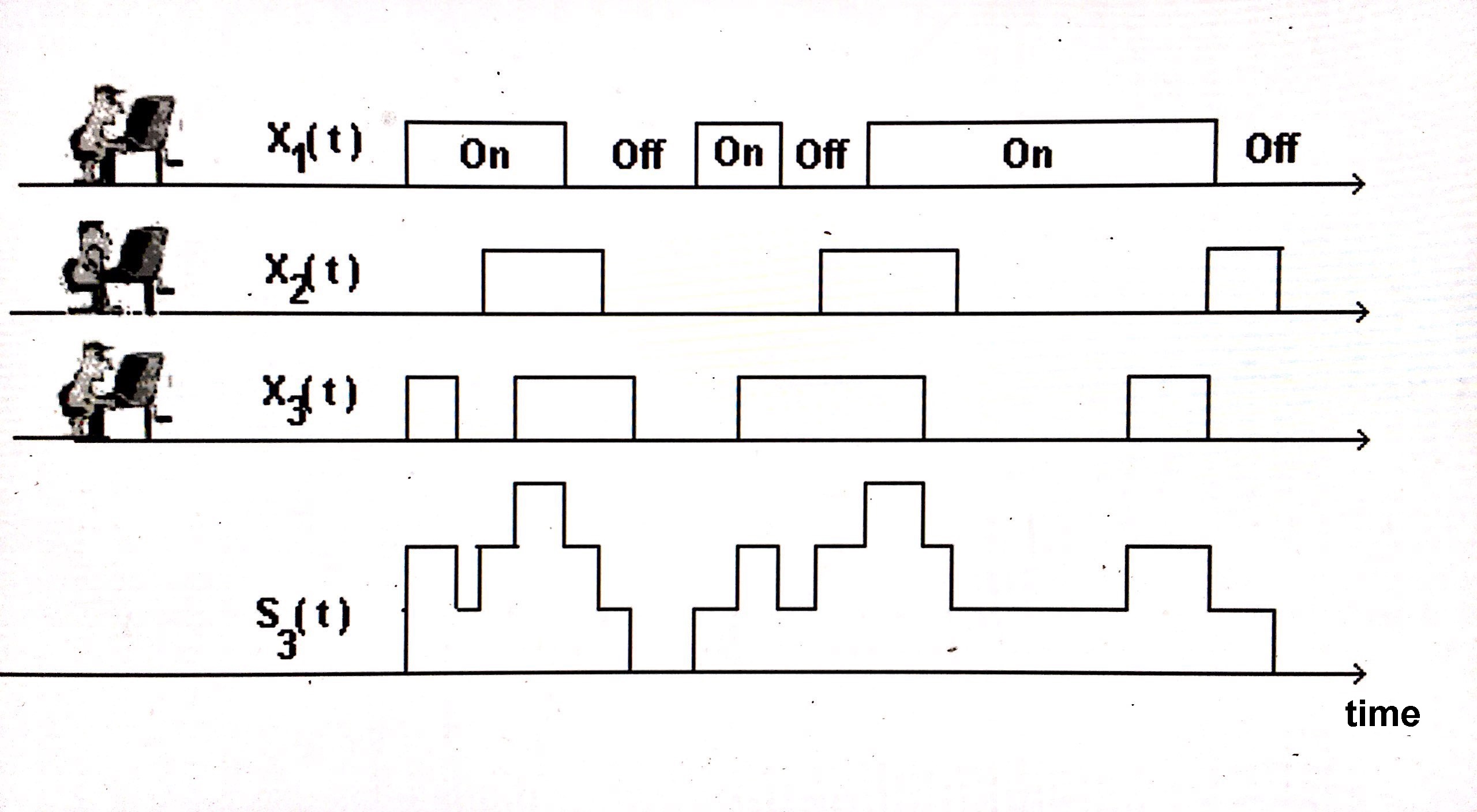}
\end{center}
\caption{Three sources \textit{On/Off}, say ($X_1(t), X_2(t), X_3(t)$), and their aggregation $S_3(t):=X_1(t)+X_2(t)+X_3(t)$.}
\label{fig:fig7}
\end{figure}

The \textit{On/Off} model considers $N$ independent sources of traffic $X_i(t), ~i\in\mathbb{N}$, where each source is a 0/1 \textit{reward-renewal} process with i.i.d periods \textit{On} and i.i.d periods \textit{Off}. This means that $X_i(t)$ assumes values 1 (\textit{On}) and 0 (\textit{Off}) alternating and non-overlapping time intervals called periods \textit{On} and \textit{Off}, respectively. Here, $X_i(t)=1$ is interpreted as being a packet transmission. Therefore a period \textit{On} can be seen as constituting a \textit{packet train}\cite{Jain_Routhier}. Three of these sources and their aggregations are shown in Figure~\ref{fig:fig7}. Since $S_N(t)$ is the process that represents the traffic aggregated at time $t$ (similar to the process $X_{nT}$ shown in Figure~\ref{fig:fig4}, so that: $S_N(t)=\sum_{i=1}^N X_i(t)$. Let us also consider the cumulative process $Y_N(Tt)$ such that $Y_N(Tt)=\int_0^{Tt} \sum_{i=1}^N X_i(s)ds$ where $T>0$ is a scale factor that is explicitly incorporated. Therefore, $Y_N(Tt)$  measures the total traffic up to time $Tt$.

Let us now consider the time distribution of \textit{On} and \textit{Off} periods. Let $\tau_{on}$ be the random variable that describes the duration of the \textit{On} and $\tau_{off}$ \textit{Off} periods. If the distribution of $\tau_{on}$ is a heavy-tailed one, $N$ and $T$ are sufficiently large, then $Y_N(Tt)$ behaves asymptotically as a FBM process \cite{Willinger_et_alli2}, \cite{Park_Kim_Crovella}, \cite{Feldmann_et_alli2}, \cite{Beran2}, \cite{Willinger_et_alli3}, \cite{Jain_Routhier}, \cite{Taqqu_et_alli}, \cite{Kihong_et_alli}, \cite{Adas_Mukherjee} \cite{Tsybakov_Georganas}, \cite{Grossglauser_Bolot}, \cite{Ryu_Elwalid}. \cite{Duffield_Whitt}, being $H=(3-\beta)/2$, where 
$\beta$ is a parameter of the long tail distribution. That is, for generate self-similar traffic and LDR is necessary $\beta$ in the interval $1<\beta<2$. In the case of Pareto distribution, briefly mentioned in Section II.A, it shows a long tail in this support $[\alpha,\infty)$ and its density function
probability can be written as $f_X(x)=\left ( \frac{\beta}{\alpha} \right )\left ( \frac{x}{\alpha} \right )^{\beta-1}.$

\begin{figure}[!ht]
\begin{center}
\includegraphics[width=3.5in]{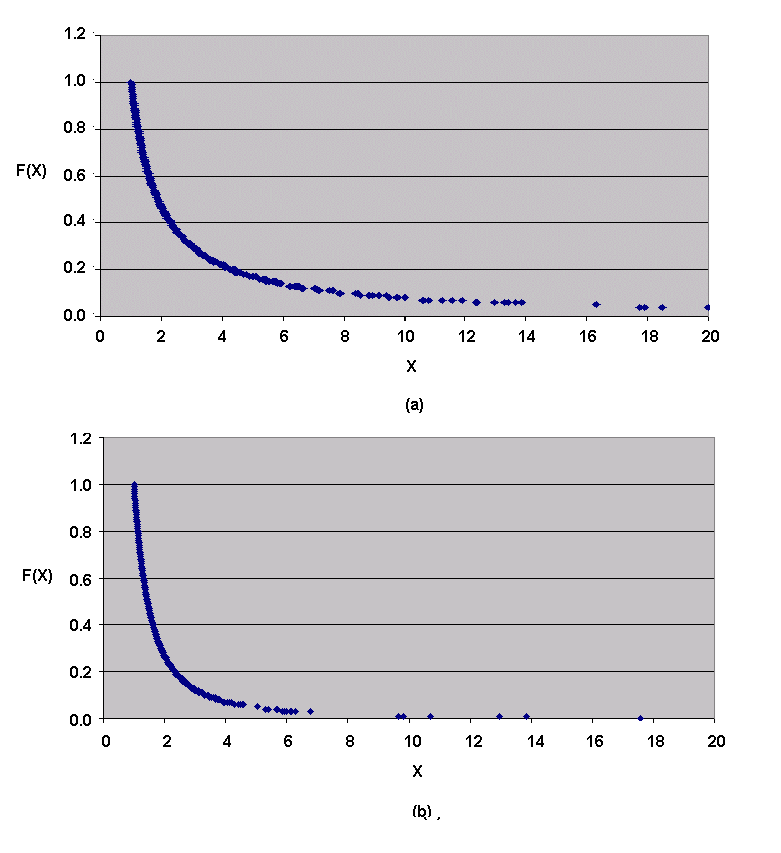}
\end{center}
\caption{Pareto probability density with parameters $\alpha, \beta$. (a) Pareto with parameters $\alpha =1$ and $\beta=1.1$; in this case a long tail is observed. (b) Pareto with parameters $\alpha=1$ and $\beta=1.9$: for this case a reduction in the tail is observed, approaching an exponential distribution that characterizes a Poisson and SRD process.}
\label{fig:fig8}
\end{figure}

Figure~\ref{fig:fig8} shows the Pareto distribution for the following parameters: Figure ~\ref{fig:fig8}(a) $\alpha=1$. $\beta$=1.1. In Figure~\ref{fig:fig8}(b) $\alpha=1$, $\beta$=1.9. It is observed that the more the parameter $\beta$ approaches 2, the distribution becomes short-tailed, approaching the exponential distribution, and leads to a process SRD, i.e., a Poisson-like process.

Different heavy tail distributions can be used to adjust the traffic model. A method for generating a large number of such distributions from known probability distributions was introduced \cite{deO_cintra}. Studying a specific network traffic scenario, an investigation would be interesting modifying the type of distribution in order to result in a better adherence to the actual data.

\subsection{Self-Similar Traffic Generator Implementation of the \textit{On/Off} Model}
\begin{figure}[!ht]
\begin{center}
\includegraphics[width=3.5in]{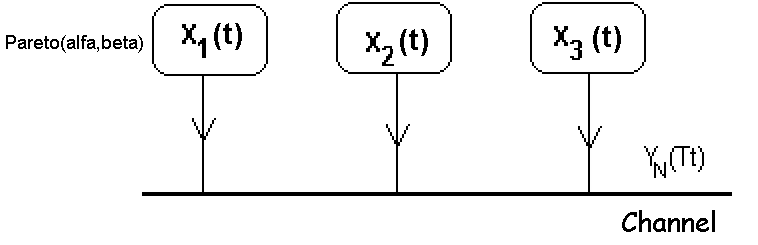}
\end{center}
\caption{Representation of aggregate traffic $Y_N(Tt)$ for $N=3$ independent sources, over a communication channel ($N=3$ just to present a short illustration). Time in which source transmits obeys a Pareto distribution with the same values of the parameters $\alpha$ and $\beta$ for \textit{On} and  \textit{Off} times and with constant transmission rate.}
\label{fig:fig9}
\end{figure}

Consider the scenario shown in Figure ~\ref{fig:fig9}. Here, $N$ independent sources of traffic transmit according to the model \textit{On/Off}, on a lossless channel, in which the traffic is added. The traffic sources are shown as closed boxes, which are associated with a transmission time that follows a long-tail distribution. In this case, it is assumed that each source has a Pareto-distributed traffic with the same parameters $\alpha$ and $\beta$ for times \textit{On} and \textit{Off}, and with constant transmission rate.
Based on the \textit{On/Off} model described in the scenario of Figure ~\ref{fig:fig9}, the following pseudocode was implemented for the generation of aggregate traffic $Y_N(Tt)$.\\
\begin{figure}[!ht]
\begin{center}
\includegraphics[width=2.5in]{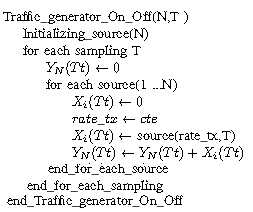}
\end{center}
\label{}
\end{figure}

The procedure Initializing\_sources ($N$): it randomly generates the initial on/off state of each source. The random generation is done by a normal distribution $\sim N(0,1)$, where 0 (Off state) is assigned to the negative values generated by $\sim N(0,1)$ and 1 (state On) to the positive values generated by $\sim N(0,1)$. 

The  procedure source($rate\_tx,~T$) returns the amount of packet transmitted from each traffic source at each sampling time T and uses the Pareto distribution for the generation of times in the \textit{On/Off} states over the sampling time $T$.

\subsection{Implementation Results of the On/Off model}

\begin{figure}[!ht]
\begin{center}
\includegraphics[width=3.5in]{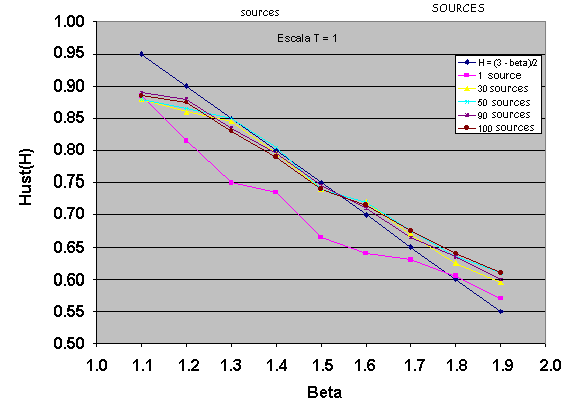}
\end{center}
\caption{Plot of the Hurst parameter $H$ as a function of the $\beta$ parameter, for the time-scale $T=1$ unit and several ensemble of sources ($N$). For the aggregate variance, the value of $\beta$ can be estimated by the slope of the straight line of the plot $\log(Variance Aggregate(m)\times \log (m)$, where $m$ is the aggregation level adopted.}
\label{fig:fig10} 
\end{figure}

For a closer result of the reference model $H=(3-\beta)/2$, several combinations of number of sources ($N$), time of sampling ($T$) and several seeds for the generation of random numbers were experimented. The best result is described in Figure ~\ref{fig:fig10}, which shows the relationship between $\beta$ and $H$, which is close to what was expected. The generated traffic had the assessment of its degree of self-similarity evaluated by the methods of aggregate variance and R/S.\\

In an Ethernet local area network model, the stations make requests to the server at short time intervals $\tau_{on}$ and spend a long time $\tau_{off}$ without transmitting, while the server has opposite behavior. In this scenario times $\tau_{on}$  of  customer sources have short tail $\beta=1.9$ and the times $\tau_{off}$  long tail $\beta=1.1$. To the server source, consider $\tau_{on}$ with long tail, $\beta=1.1$ and $\tau_{off}$ short tail with $\beta=1.9$.

\begin{figure}[!ht]
\begin{center}
\includegraphics[width=3.5in]{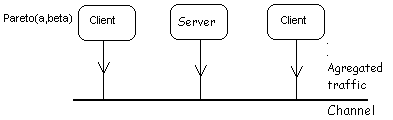}
\end{center}
\caption{Scenario of a client/server packet network with client-side traffic sources with short tail distribution transmission time (Pareto $\beta= 1.9$) and with a single source of server traffic with long tail transmission time distribution (Pareto  $\beta=1.1$).}
\label{fig:fig11a}
\end{figure}

\begin{figure}[!ht]
\begin{center}
\includegraphics[width=3.5in]{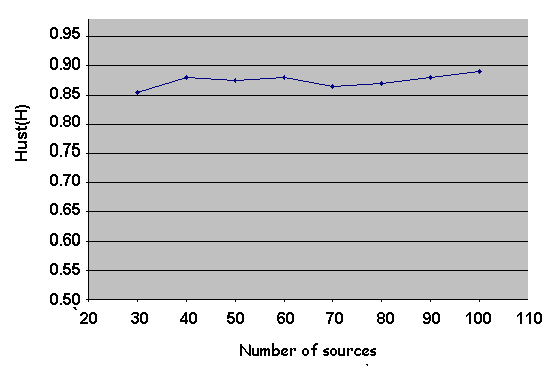}
\end{center}
\caption{Variation of self-similarity degree ($H$) as a function of the number of sources for the scenario shown in Fig. ~\ref{fig:fig11a}.}
\label{fig:fig11b}
\end{figure}

Figure ~\ref{fig:fig11a} shows the scenario described in the previous paragraph. Figure ~\ref{fig:fig11b} shows the influence of the number of \textit{On/Off} sources (as described in the previous paragraph) on the Hurst parameter. These results are in agreement with the traffic generation for a self-similar model described in \cite{Willinger_et_alli1}, which suggests that traffic in a Ethernet network may have a self-similar behavior, in the case with degree of self-similarity around 0.85. 

As a remark, Figure \ref{fig:fig11b} suggests that in a network of client and server packages with \textit{On-Off} time characteristics similar to those considered, the traffic may have an auto-similar behavior, in the case, with a degree of self-similarity around 0.85. If these results are compared to those obtained in measurements made on Ethernet LAN in \cite{Leland_et_alli} ($H=0.80$) and modeling and traffic generation \textit{On/Off} \cite {Kihong_et_alli} ($H=0.90$), and our results can be considered as  satisfactory.
\subsection{Implications of self-similarity on network performance}

\begin{figure}[!ht]
\begin{center}
\includegraphics[width=3.5in]{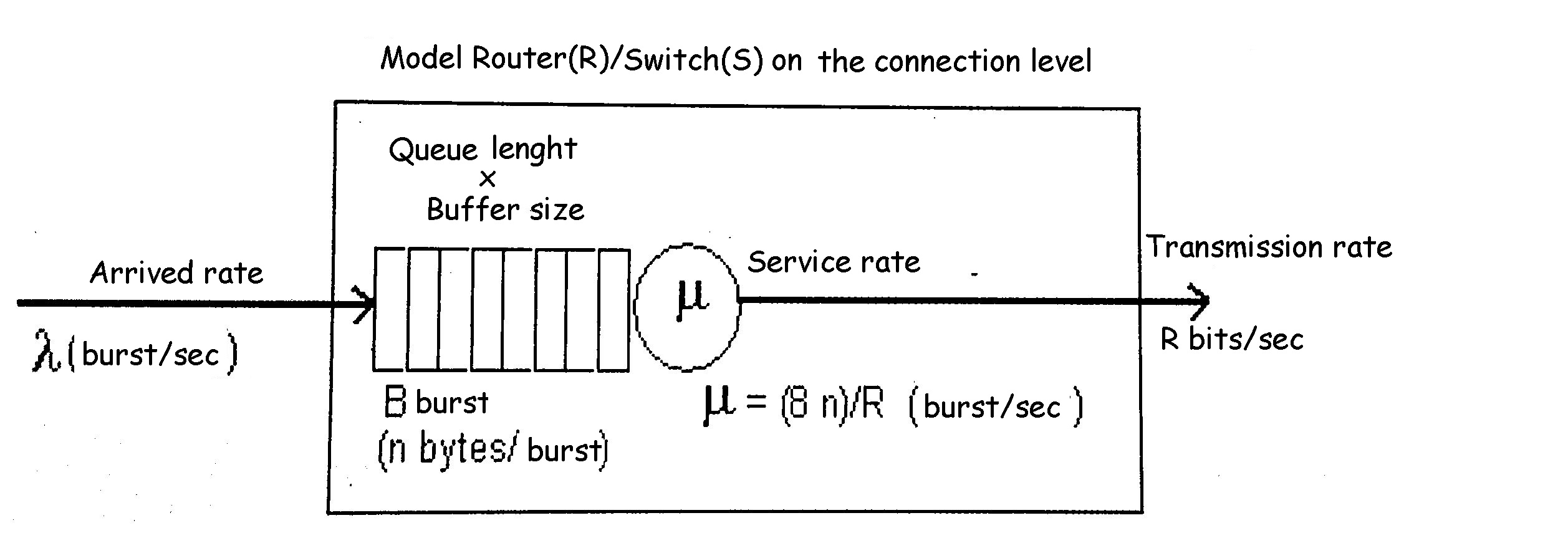}
\end{center}
\caption{Representation of a contention element model at the connection level.}
\label{fig:fig13}
\end{figure}

\begin{figure}[!ht]
\begin{center}
\includegraphics[width=3.5in]{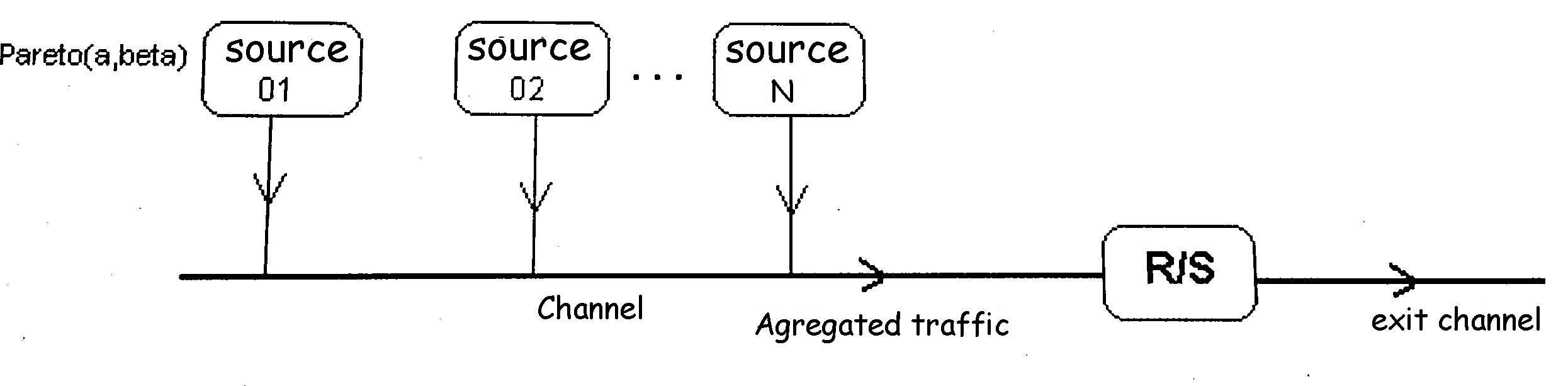}
\end{center}
\caption{Representation of a contention element in a network framework.}
\label{fig:fig14}
\end{figure}

\begin{figure}[!ht]
\begin{center}
\includegraphics[width=3.5in]{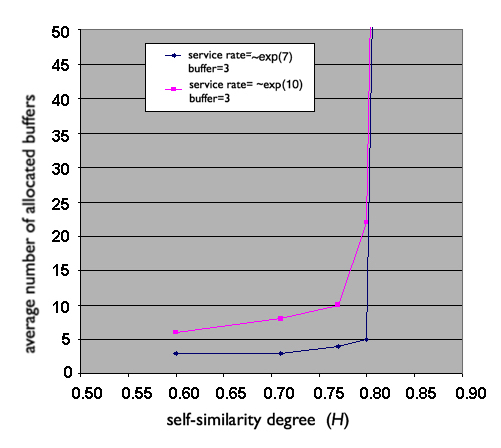}
\end{center}
\caption{Average number of buffers according to the degree of self-similarity for arrival traffic in the R/S contention element, the mean number increases steeply when the degree of self-similarity of the incoming traffic approaches 0.80. The same behavior is observed when the service rate is constant or variable with an exponential distribution with mean $\mu$.}
\label{fig:fig15}
\end{figure}

\begin{figure}[!ht]
\begin{center}
\includegraphics[width=3.5in]{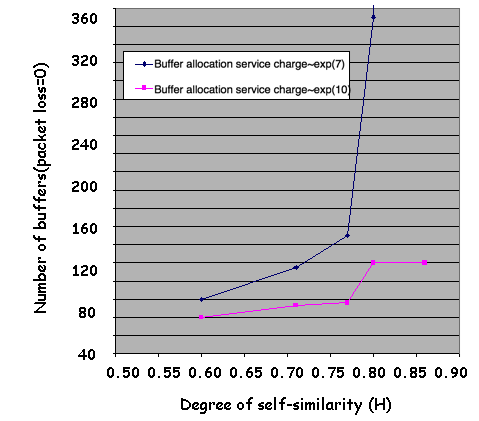}
\end{center}
\caption{Number of buffers for a packet loss (zero overflow), based on the degree of self-similarity of the incoming traffic. In the allocation strategy of buffers a  service rate variable exponential distribution with mean $\mu=7$ and a buffer width=3, the number of required buffers grows abruptly to a degree of self-similarity close to 0.80. For the service rate increase strategy from 7 to 10, the number of required buffers remains low.}
\label{fig:fig16}
\end{figure}

The performance of a network is directly related to the performance of the R/S containment elements (Router/Switch). Figure~\ref{fig:fig13} illustrates a typical containment element, in this case, the routing element \cite{McDysan}. This element presents two processes, one of arrival and one of service. Figure ~\ref{fig:fig14} illustrates the context of this contention element in a network. Several scenarios of arrival self-similar and Poisson service were simulated. The simulation results are shown in Figure~\ref{fig:fig15}, allocation of buffers as a function of degree of self-similarity of the traffic generated by the sources, and the Figure 16, allocation of buffers to avoid packet loss due to variation of the degree of self-similarity. It can be observed in Figure ~\ref{fig:fig15} that the allocation of buffers grows abruptly when the degree of self-similarity increases, the same occurs with packet loss in Figure~\ref{fig:fig16}. Buffers limitation lead to discard of packets and  network degradation.

It is observed in Figure~\ref{fig:fig15} that the allocation of buffers grows abruptly when the degree of self-similarity increases, the same occurs with packet loss in Figure~\ref{fig:fig16}. Buffers limitation lead to discard of packets and degradation of the network.

The strategy for reducing packet loss based on increasing buffers seems to lead to an excessive increase in queue size and  consequent packet send  delay. The strategy of increasing the service rate with the increase of the output bandwidth seems to be more suitable because it reduces packet loss and maintains the  queue size small, thus reducing the delay in the packet dispatch. However, this strategy may lead to under utilization or waste of the channel band, due to the variable characteristic of self-similar traffic.

\section{Conclusions}

Self-similarity in the traffic of computer networks is not at all an up-to-the-minute discovery and neither is the work in this area, even though today's networks are dominated by fiber optic technology, Ethernet and TCP/IP stack protocols, which reinforce the phenomenon. However, a few researchers and network designers insist on using the Poisson processes to generate the traffic model, which, as argued in this paper, does not reflect nor approximate the actual behavior of the present-day networks, often leading to the incorrect dimensioning of resources (mainly of contention elements and channel bandwidth). In this tutorial, the need for adoption of models that take into account the self-similarity present in traffic is emphasised. Heavy-tailed distributions are nice tools to build a suitable model. By studying a specific network traffic scenario, an investigation would be interesting by modifying the heavy-tailed distribution so as to result in better adherence to the data.

\section*{Acknowledgment}
The authors particularly thank the Editor of RTIC, Dr. Wamberto Queiroz, and an anonymous and careful reviewer who helped to improve this article in many detail.

%



\ifCLASSOPTIONcaptionsoff
  \newpage
\fi

\end{document}